\documentclass[twocolumn,showpacs,preprintnumbers,amsmath,amssymb]{revtex4}
\usepackage{tabularx,graphicx}\begin{document}
\newcommand{\beq}{\begin{equation}}
\newcommand{\eeq}{\end{equation}}
\newcommand{\beqn}{\begin{eqnarray}}
\newcommand{\eeqn}{\end{eqnarray}}
\newcommand{\bmath}{\begin{subequations}}
\newcommand{\emath}{\end{subequations}}
\newcommand{\bra}[1]{\langle #1|}
\newcommand{\ket}[1]{|#1\rangle}

\title{Kinetic energy driven superfluidity and superconductivity and the origin of the Meissner effect}
\author{J. E. Hirsch }
\address{Department of Physics, University of California, San Diego,
La Jolla, CA 92093-0319}

\begin{abstract} 
Superfluidity and superconductivity have many elements in common. However, I argue that their most important commonality has been overlooked: that both are kinetic energy driven. Clear evidence that superfluidity in $^4He$ is kinetic energy driven is the shape of the $\lambda$ transition and the negative thermal expansion coefficient below $T_\lambda$. Clear evidence that superconductivity is kinetic energy driven is the Meissner effect: I argue that otherwise the Meissner effect would not take place.
Associated with this physics I predict that superconductors expel negative charge from the interior to the surface and that a spin current exists in the ground state of superconductors (spin Meissner effect). I propose that this common physics of superconductors and superfluids originates in rotational zero point motion. This view of superconductivity and superfluidity implies that rotational zero-point motion is a fundamental property of the quantum world that is missed in the current understanding. \end{abstract}
\maketitle 

\section{introduction}

   \begin{figure}
 \resizebox{8.5cm}{!}{\includegraphics[width=9cm]{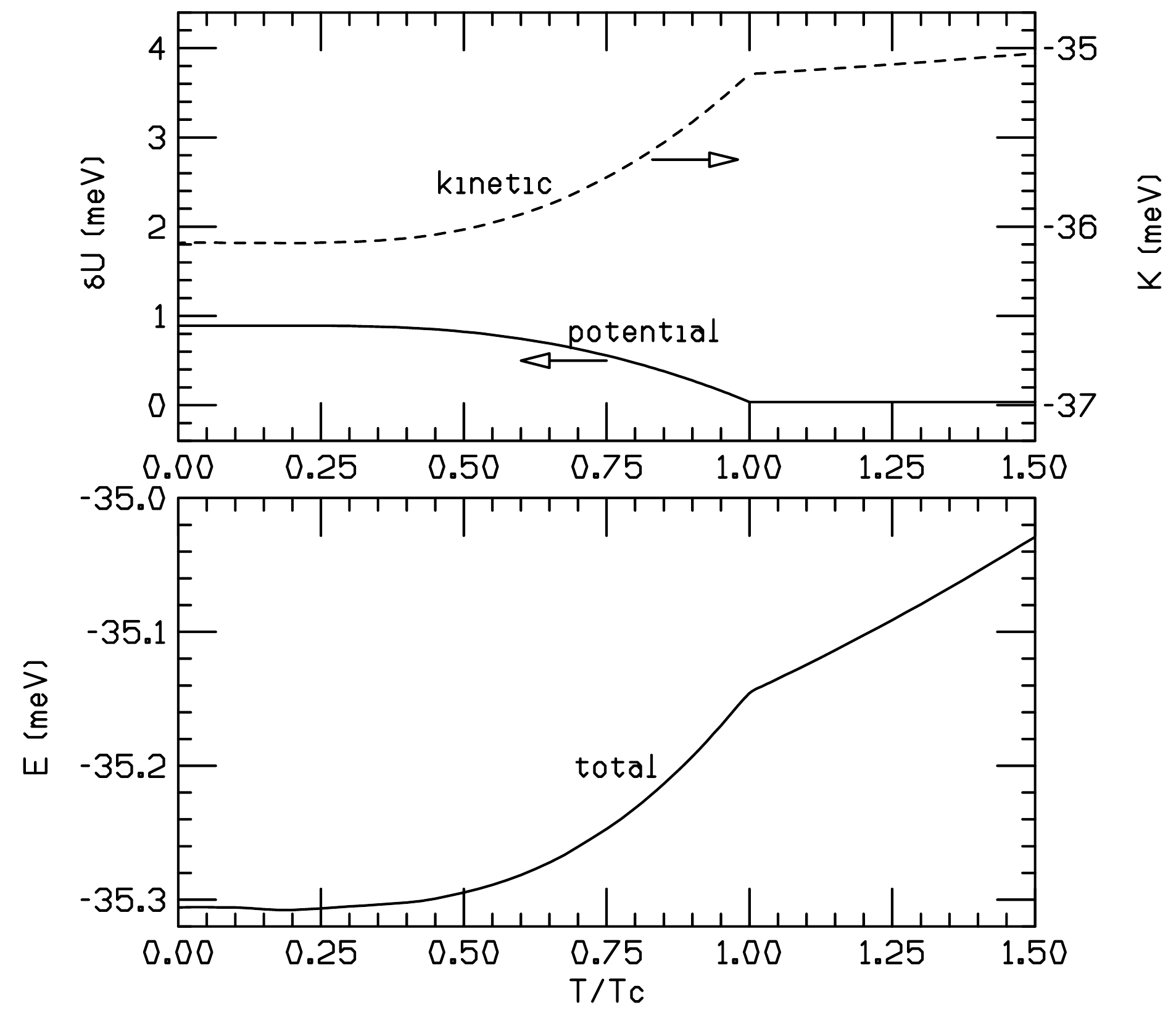}}
 \caption {Kinetic ($K$), potential ($\delta U$) and total ($E$) energies  in $meV$ per atom
 versus reduced temperature in the model of hole superconductivity. The model parameters correspond to the
 case of Ref. \cite{99} figure 1, for hole concentration $n=0.045$ corresponding to a  $T_c$ of $85K$. 
 For the potential energy ($\delta U$) the Hartree contribution $Un^2/4$, with $U$ the on-site Coulomb repulsion, which is independent of temperature, has been substracted.
 }
 \label{figure2}
 \end{figure}

    \begin{figure}
 \resizebox{8.5cm}{!}{\includegraphics[width=9cm]{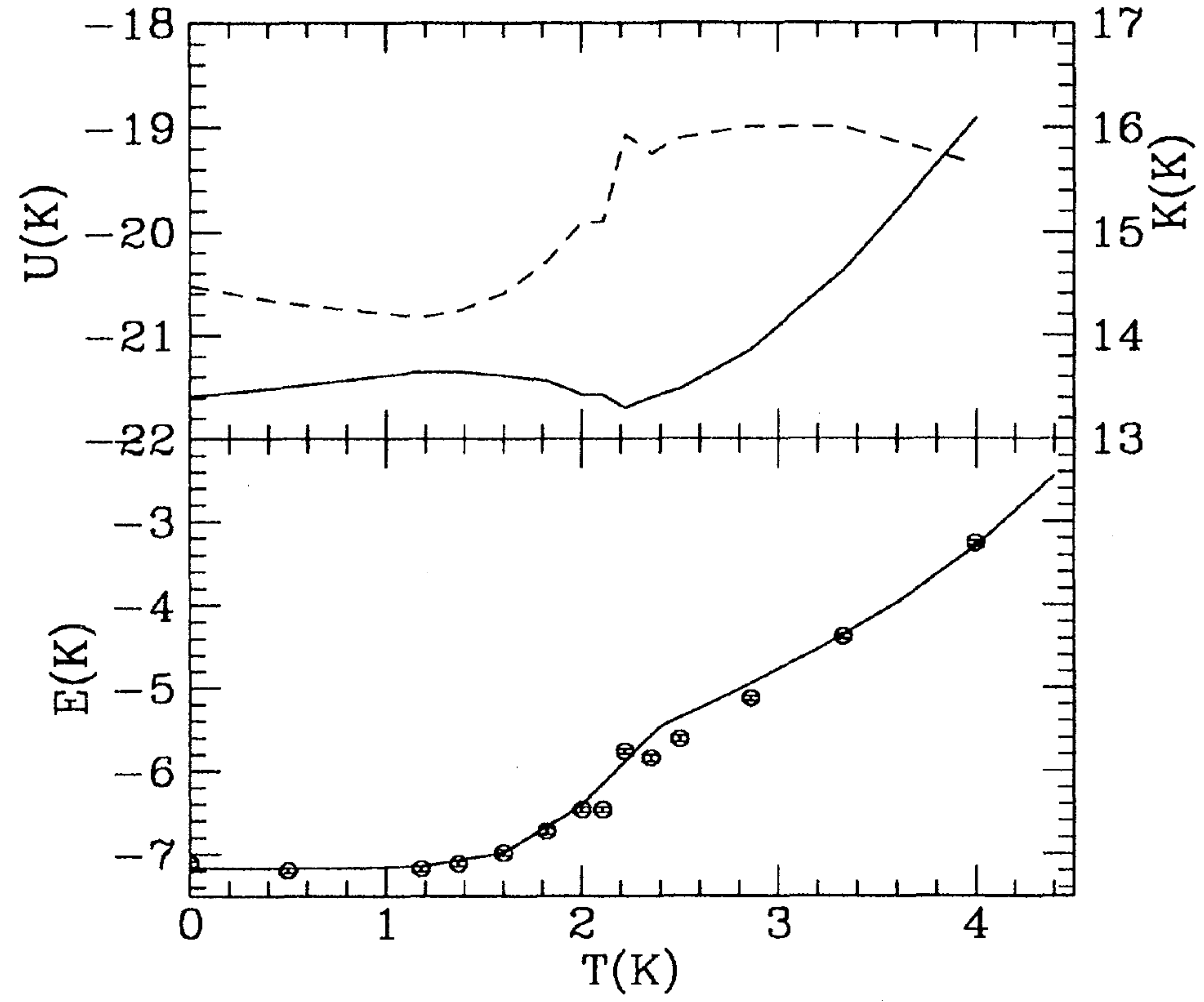}}
 \caption {Kinetic ($K$, dashed line upper panel), potential ($U$, full line upper panel) and total ($E$, lower panel) energies  in $^o K$ per $^4He$ atom versus  temperature 
computed using path integral Monte Carlo by D. Ceperley\cite{ceperley}. The points in the lower panel are experimental data, see Ref. \cite{ceperley}.
 }
 \label{figure2}
 \end{figure}
 
       \begin{figure}
 \resizebox{8.5cm}{!}{\includegraphics[width=9cm]{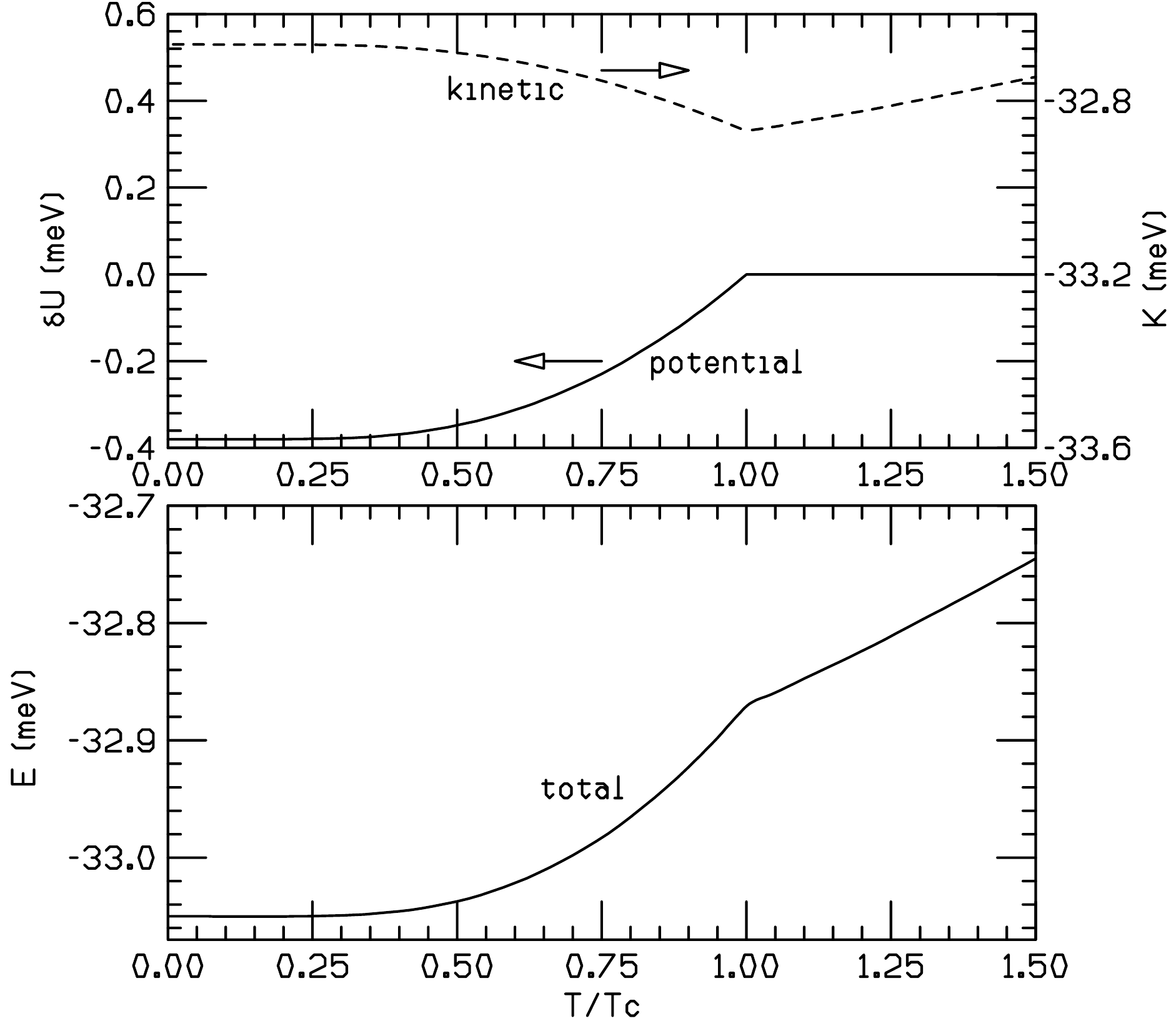}}
 \caption {Same as Fig. 1 for an attractive Hubbard model representative of conventional BCS. The $T_c$ and band filling are  the same as in Fig. 1. $U=-0.4$.
  }
 \label{figure2}
 \end{figure}
 
That superconductivity and superfluidity   have many common elements is certainly well known\cite{london12,tilley}. An indication of this is
that  the terms ``superfluid electrons''
and ``superfluid condensate'' are commonly used to refer to the charge carriers in the superconducting state of a metal. However I propose that a deep commonality between 
superconductors and superfluid $^4He$  has been overlooked until now: that both phenomena are
{\it kinetic energy driven}.

Figures 1 and 2 show kinetic, potential and total energies versus temperature for the model of hole superconductivity\cite{99} and for superfluid $^4He$ computed
through Monte Carlo simulations  by 
D. Ceperley\cite{ceperley} (direct experimental data on kinetic and potential energies separately do not exist). The similarity
in the two figures is very apparent. The potential energy increases 
as the system enters the superfluid or superconducting state, while the kinetic energy decreases, hence the ``super'' state is
``kinetic energy driven'' in both cases. 

In contrast, within conventional BCS theory  the kinetic energy of the carriers always  increases upon entering the superconducting state and the interaction
energy decreases by a larger amount overcompensating the kinetic energy increase, as shown in Figure 3, 
hence superconductivity is ``potential energy driven''.   The physics displayed in Figure 3 is  $qualitatively$ different from the physics
shown in Figures 1 and 2. I argue that the Meissner effect results from the physics shown in Fig. 1 and would not occur if the physics was as in Fig. 3, for reasons explained below.

That superfluidity in $^4He$ is kinetic energy driven is clear from a variety of experimental data that we will review in the next section. 
That superconductivity is kinetic energy driven is predicted by the model of hole superconductivity, introduced  in 1989\cite{holesc}. 
The pairing interaction was denoted by $\Delta t$ to indicate its kinetic origin, and its effect on the kinetic energy was discussed in Ref. \cite{apparent}.  However it was only much later that the 
fundamental physics of kinetic energy lowering, which is completely analogous to the physics taking place in $^4He$,  and its role in the Meissner effect, was understood in this model.

     \begin{figure}
 \resizebox{8.5cm}{!}{\includegraphics[width=9cm]{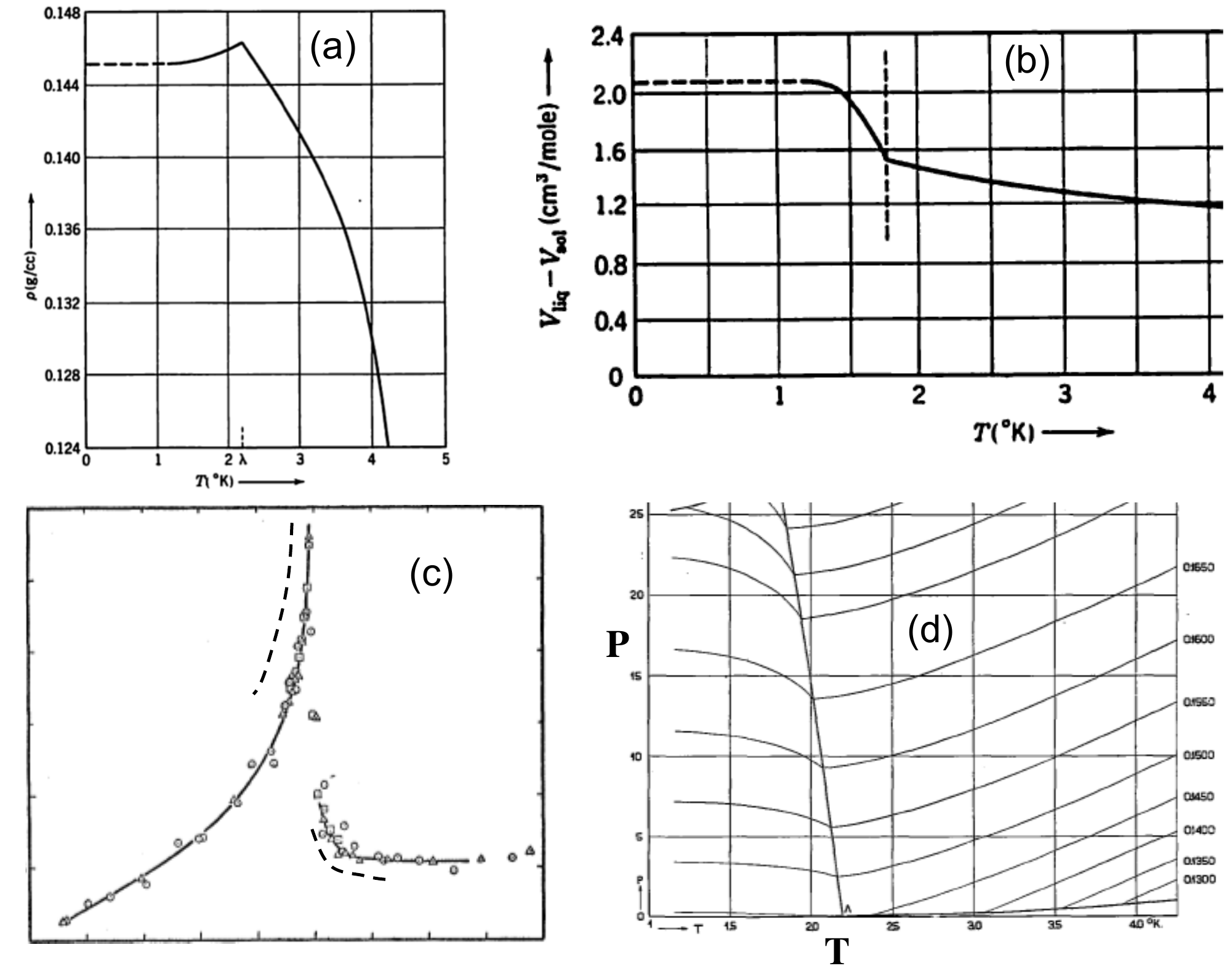}}
 \caption {Four experimental properties of $^4He$. (a) Density versus temperature at constant pressure. (b) Difference between liquid and solid molar volumes
 at the liquid-solid transition as function temperature. (c) Heat capacity versus temperature at constant volume. 
 The dashed lines show schematically the contribution of kinetic energy only to the total heat capacity. (d) Pressure versus temperature at
 constant density (isopycnals). }
 \label{figure2}
 \end{figure}
 
 \section{superfluid $^4He$ and wavefunction expansion} 
 Figure 4 shows four properties of $^4He$ that illustrate the physics of interest here. (a) shows the density versus temperature at constant pressure.
 Below the superfluid transition, there is a slight $decrease$ in the density, which is clearly $not$ driven by potential energy: the $^4He$ atoms are spherical, so there
 is no directionality to the interatomic forces, and the average distance between atoms in the liquid is $4 \AA$, while the minimum in the potential energy curve between
 $He$ atoms is at distance $3\AA$\cite{londonbook}. If the density decreases, the interatomic distance increases and the potential energy increases. Hence the decrease in
 density seen below the $\lambda$ point has to be associated with lowering of kinetic energy, i.e. is  kinetic energy driven.
 We can think of the $^4He$ atoms as being confined in a box of size determined by the interatomic distance. The kinetic energy of quantum confinement will
 decrease when the density decreases and the interatomic distances increase. 
 
 Similarly Figure 4(b) shows the increase in volume as $^4He$ goes from
 the solid to the liquid state. It becomes markedly larger at temperatures below the superfluid transition. At low temperatures the entropy of both states is zero\cite{londonbook}, so the expansion is not entropy-driven as in an ordinary solid-liquid transition 
 but energy-driven. Once again, since the potential energy increases upon expansion and the total energy decreases in going from the solid to the superfluid state this is direct evidence that the transition
 from the solid  into the superfluid state is
 kinetic-energy driven.
 
 Figure 4(c) shows the heat capacity versus temperature, the characteristic shape that gives the $\lambda$ transition its name (it should really be
 called `inverted lambda' transition. The heat capacity is given by
 \beq
 C=\frac{d<K>}{dT}+\frac{d<U>}{dT}
 \eeq
 with $<K>$ and $<U>$ the average kinetic and potential energies. The second term in this equation is positive above 
 $T_\lambda$, since the system expands as $T$ increases and hence the potential energy increases, and is negative below $T_\lambda$ since the
 system expands as $T$ decreases. Thus, the first term in Eq. (1) is even larger below $T_\lambda$ and even smaller above
 $T_\lambda$ than the full line in  Fig. 4(c) shows \cite{goldstein}, as indicated by
 the dashed lines in Fig. 4(c),
 hence the jump at $T_\lambda$ for the change in kinetic energy with $T$ is even larger. The fact that the rate of decrease of the kinetic energy as the temperature is lowered is so much larger below $T_\lambda$ than above $T_\lambda$
is clear evidence that the transition from the normal liquid into the superfluid state is kinetic energy driven\cite{goldstein}.
 
Finally, Figure 4(d) shows the pressure versus temperature at constant density\cite{isopyc}. Below $T_\lambda$, the pressure $increases$ as the temperature is lowered.
 This is qualitatively different from what occurs in ordinary Bose condensation: in that case, the condensate exerts $no$ pressure, hence
  the pressure decreases rapidly as the temperature is lowered and the condensate fraction increases. In $^4He$ instead, the pressure increases as the condensate
  forms, indicating that it exerts more   quantum pressure  than the normal fluid, causing the liquid to expand.
 
 This physics of $^4He$ is qualitatively different from Bose condensation physics. In a Bose gas, increasing the external pressure and hence the density at a fixed temperature will eventually lead to
 Bose condensation as the interatomic distances become of the order of the de Broglie wavelength. Instead, in $^4He$, 
 increasing the pressure and density at fixed temperature will  $never$ lead from the normal liquid into the superfluid state, nor from the solid into the
 superfluid state. The superfluid transition involves $expansion$, hence application of pressure or increase in density can only lead $out$ of the superfluid state (either into the solid or into 
 the normal fluid state), never $into$ it. 
 This is clearly seen in the phase diagram of $^4He$. 
 
       \begin{figure}
 \resizebox{8.5cm}{!}{\includegraphics[width=9cm]{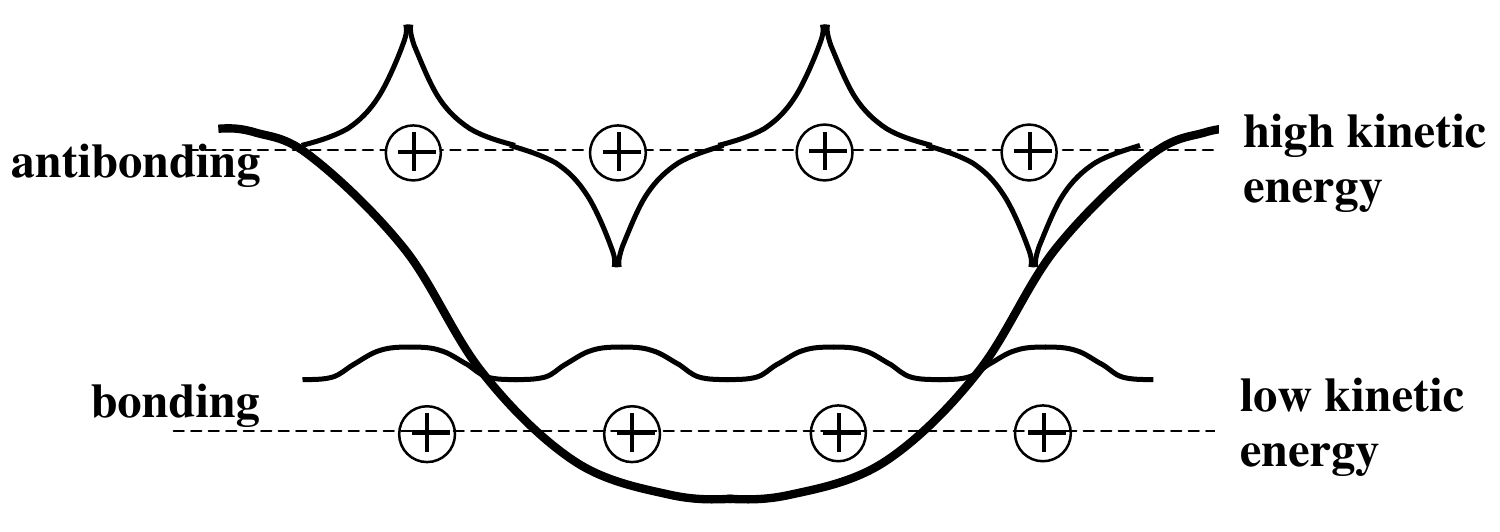}}
 \caption {Electronic states in an energy band. The states near the top of the band (antibonding states) have high kinetic energy and short wavelength. Electrons in those states
 exert strong quantum pressure outward and tend to break the lattice apart. In superconducting materials those states are occupied by electrons (the Fermi energy is close to 
 the top of the band), in non-superconductors they are empty according to the theory of
 hole superconductivity.}
 \label{figure2}
 \end{figure}
 
 The  properties of $^4He$ just summarized indicate that the transition into the superfluid state is associated with wavefunction expansion,
kinetic energy lowering and enhanced quantum pressure originating in   quantum zero-point motion\cite{simon}. 
We propose that exactly the same is true for superconductors and that this is the physics responsible for the Meissner effect.
 
  \section{Hole superconductivity and wavefunction expansion}
The theory of hole superconductivity predicts that superconductivity occurs when electronic energy bands are almost full, hence the carriers in the normal state are holes. 
When a band is almost full, there are a lot of {\it antibonding electrons}, as shown schematically in Fig. 5. They would like to break the solid apart, hence their name, ``antibonding''.
Their wavefunction is confined over a small spatial dimension, their wavelength $k_F^{-1}$ is short ($k_F$ is the Fermi wavevector), 
and they exert ``quantum pressure'' outward. They have highly oscillating wavefunctions and hence high kinetic energy.

  \begin{figure}
 \resizebox{8.5cm}{!}{\includegraphics[width=9cm]{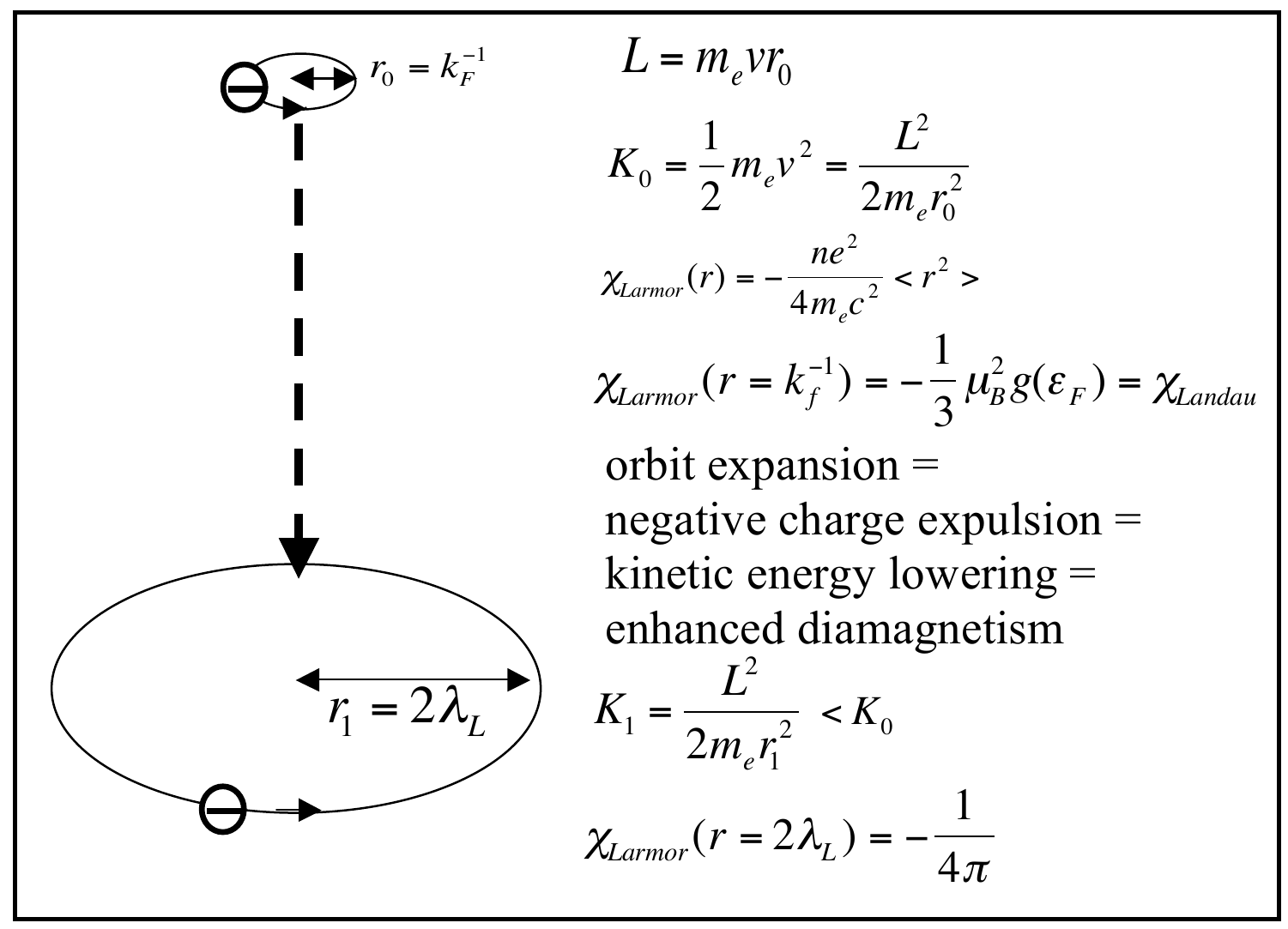}}
 \caption {Explanation of the Meissner effect. An electron in an expanding orbit with fixed angular momentum
 lowers its kinetic energy ($K_1<K_0$), increases its Larmor diamagnetic susceptibility and causes expulsion of negative charge.
 The top orbit represents the normal state, with $r_0=k_F^{-1}$, the bottom one the 
 superconducting state, with $r_1=2\lambda_L$.}
 \label{figure2}
 \end{figure}

Within the theory of hole superconductivity\cite{website}, pairing of holes occurs at the critical temperature because it gives rise to kinetic energy lowering\cite{holesc,apparent}. When holes pair, the band becomes locally less full, hence the kinetic energy
should decrease according to Figure 5. In addition, the pairing interaction $\Delta t$ gives rise to kinetic energy lowering for the pair. 
The transition
to superconductivity is associated with expansion of the electronic wavefunction and expulsion of negative
charge from the interior of the superconductor to a region within a London penetration depth of the surface, $\lambda_L$\cite{electronhole2,chargeexpulsion}. 
The expansion of the wavefunction and negative charge expulsion results from an expansion of electronic orbits from microscopic radius $k_F^{-1}$ 
 to mesoscopic radius $2\lambda_L$\cite{sm}, which lowers the quantum kinetic energy, and changes the diamagnetic susceptibiliy from the Landau free electron value to the value appropriate for perfect
diamagnetism, $\chi=-1/4\pi$,  as shown schematically in Figure 6. The expansion of electronic orbits and associated outward motion of negative charge  provides a dynamical
explanation of the Meissner effect\cite{meissner}: in the presence of a magnetic field, the Lorentz force on the radially outgoing electrons deflects them in the azimuthal direction giving rise to the
Meissner current that expels the magnetic field from the interior. In other words, the outflowing charge carries with it the magnetic field lines, as in a classical plasma\cite{plasma}.
Instead, if there is no radial motion of charge, as expected within BCS theory, magnetic field lines would not move out, there would be no Meissner effect, and the material would not
become a superconductor\cite{japan}.

 The fact that superfluid electrons in the superconducting state reside in orbits of radius $2\lambda_L$ can  be seen from the fact that 
 the total angular momentum of electrons in  such orbits  equals the angular
momentum of the Meissner current circulating within a London penetration depth of the surface in a cylindrical geometry, as shown by the
following equation:
\beq
L_{total}=[m_e v (2\lambda_L)]n_s[ \pi R^2 h]=[m_e v R] n_s [2\pi R \lambda_L]
\eeq
where $R$ and $h$ are the radius and height of the cylinder and $n_s$ is the superfluid density. 
      \begin{figure}
 \resizebox{8.5cm}{!}{\includegraphics[width=9cm]{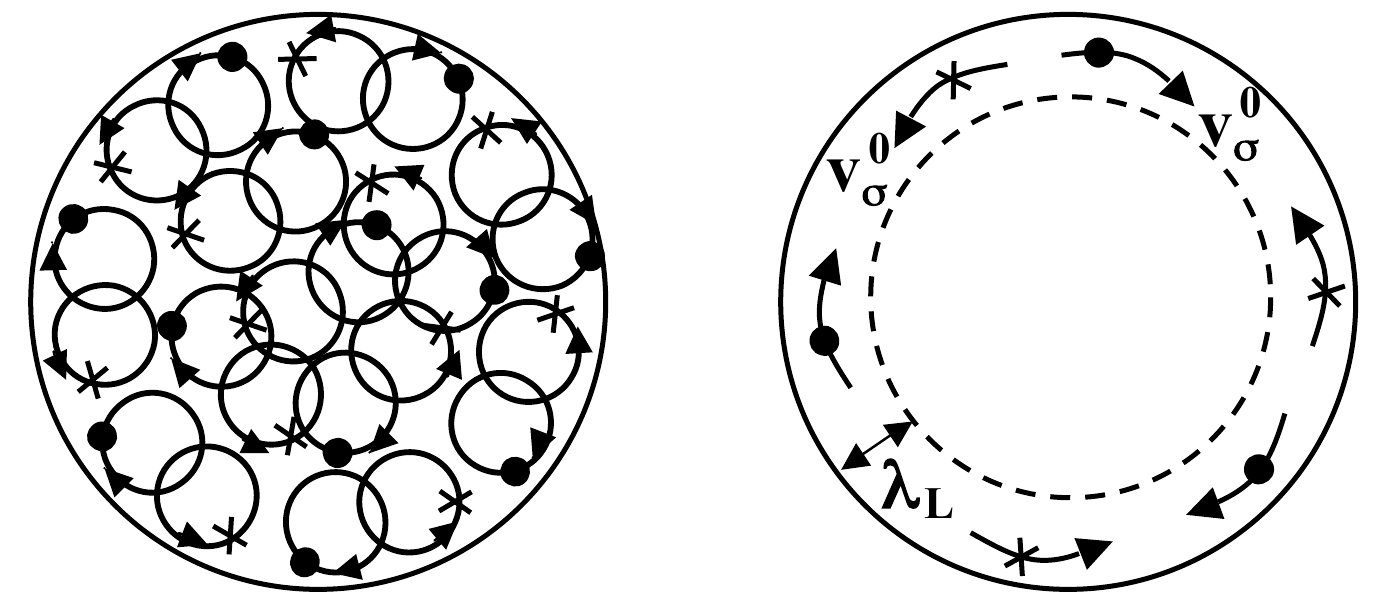}}
 \caption { The left side shows electronic orbits of radius $2\lambda_L$, with electrons with spin pointing into the paper (out of the paper) circulating in counterclockwise (clockwise) direction.
 The orbits are highly overlapping.
 The superposition of these motions  (right side)
  gives rise to a spin current circulating in a layer of thickness $\lambda_L$ near the surface in the ground state of the superconductor, and no net currents in the interior.}
 \label{figure2}
 \end{figure}
Electrons in the $2\lambda_L$ orbits traverse these   orbits with speed given by\cite{sm}
\beq
v_\sigma^0= \frac{\hbar}{4m_e\lambda_L} .
\eeq
in opposite direction for opposite spin. The superposition of these motions gives rise to a macroscopic spin current of carrier density $n_s/2$ for each spin direction  flowing within a London penetration depth of the
surface with speed Eq. (3),   a macroscopic
zero point motion of the superfluid.\cite{electrospin} This is shown schematically in Figure 7. 

As a result of this orbit expansion, the electronic density in the interior of the superconductor is
slightly smaller than in the normal state. This is entirely analogous to the density decrease that occurs in $^4He$ upon the onset of superfluidity.   
The excess negative charge near the surface has density $\rho_-$, related to the speed of the spin current Eq. (3)  through the equation\cite{electrospin}
 \beq
 \rho_-=en_s\frac{v_\sigma^0}{c}  .
 \eeq
 Thus, we can think equivalently of the entire superfluid charge density $en_s$  flowing with speed Eq. (3) (half in each direction) or just the excess charge density $\rho_-$ flowing at the speed of light.
 The orbital angular momentum of superfluid electrons in the $2\lambda_L$ orbits is
 \beq
 L_{orb}=m_e v_\sigma^0 (2\lambda_L) =\hbar/2 .
 \eeq
 
        \begin{figure}
 \resizebox{8.5cm}{!}{\includegraphics[width=9cm]{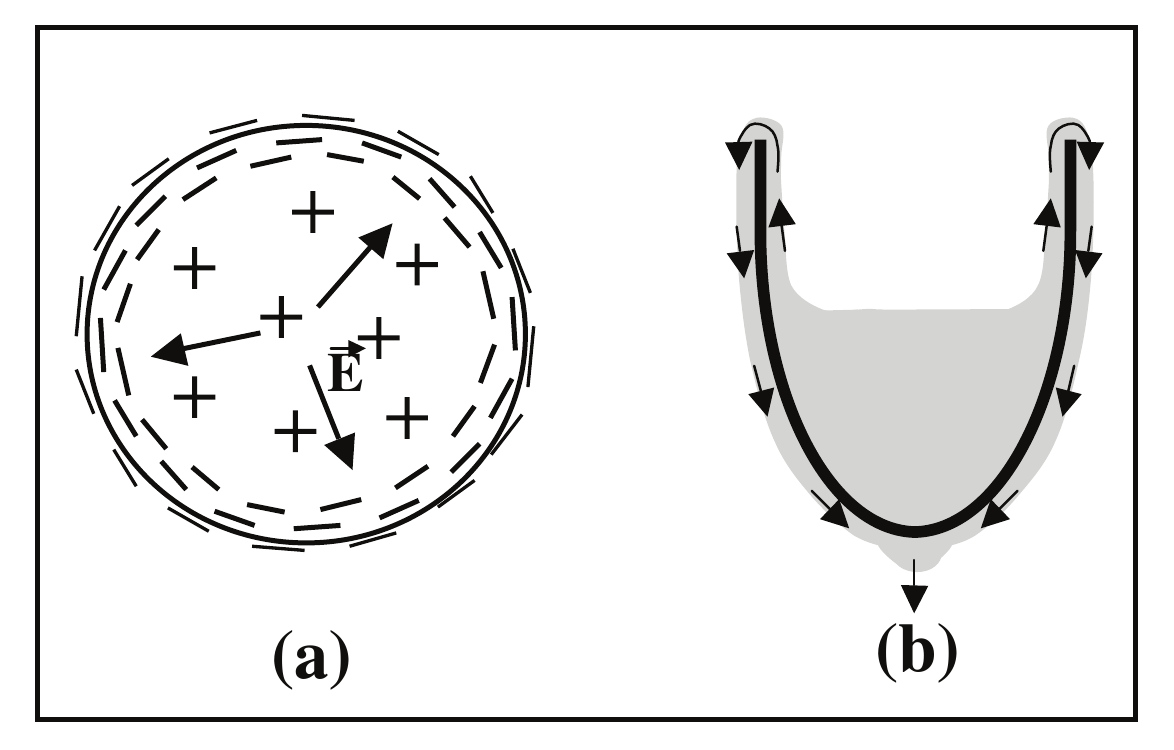}}
 \caption { The expulsion of charge from the interior of the superconductor (a) has as counterpart the expulsion of mass from the superfluid $^4He$ container (b), climbing the lateral surfaces and escaping to 
 the exterior   (``Onnes effect'').}
 \label{figure2}
 \end{figure}
 
The question arises whether the electronic orbit expansion will give rise to a lower density for the solid as a whole when it becomes superconducting. This is indeed seen in many superconductors\cite{thexp1,thexp2} but not 
in all. The situation is more complicated than in $^4He$ because of the presence of electronic and ionic degrees of freedom.

 \section{zero point motion in superfluid $^4He$ and  in superconductors}
 
 The fact that we have found charge expulsion and macroscopic zero point motion in the superconductor, resulting from expansion of the electronic wavefunction, suggests that  similar effects should occur in superfluid $^4He$.
 Remarkably, such behavior has been known for a long time: the `Onnes effect'\cite{onnes},  the flow of superfluid films (Rollin films)\cite{rollin} along surfaces 
without any driving force\cite{films}.  A superfluid container will expel mass, just like the 
 superconductor expels charge, as shown schematically in Figure 8.  $^4He$ atoms flow in the Rollin film defying the force of gravity, just as electrons develop the Meissner current defying the
 Faraday electromotive force\cite{emf}.

  The close connection between superconductors and superfluid $^4He$ becomes even more apparent when we consider superfluid flow under zero potential difference.
 This is achieved in a superconducting wire inserted between normal conductors, and in the $^4He$ double beaker experiment of Daunt and Mendelssohn\cite{beaker}, designed specifically for this purpose, as
 shown schematically in Figure 9. 
  Mendelssohn\cite{mendel,mendel2} pointed out the clear analogy between the phenomena shown in Figures 9 (a) and (b) and asked the question, what is the dynamical origin of these motions that occur without
 potential drop, i.e. without a force? He proposed that they are evidence for {\it zero point motion} of the condensed particles in the superfluid and in the superconductor. He points out that
 ``neither case corresponds to a Bose-Einstein condensation since both have an appreciable zero-point energy''.

   \begin{figure}
\resizebox{8.5cm}{!}{\includegraphics[width=7cm]{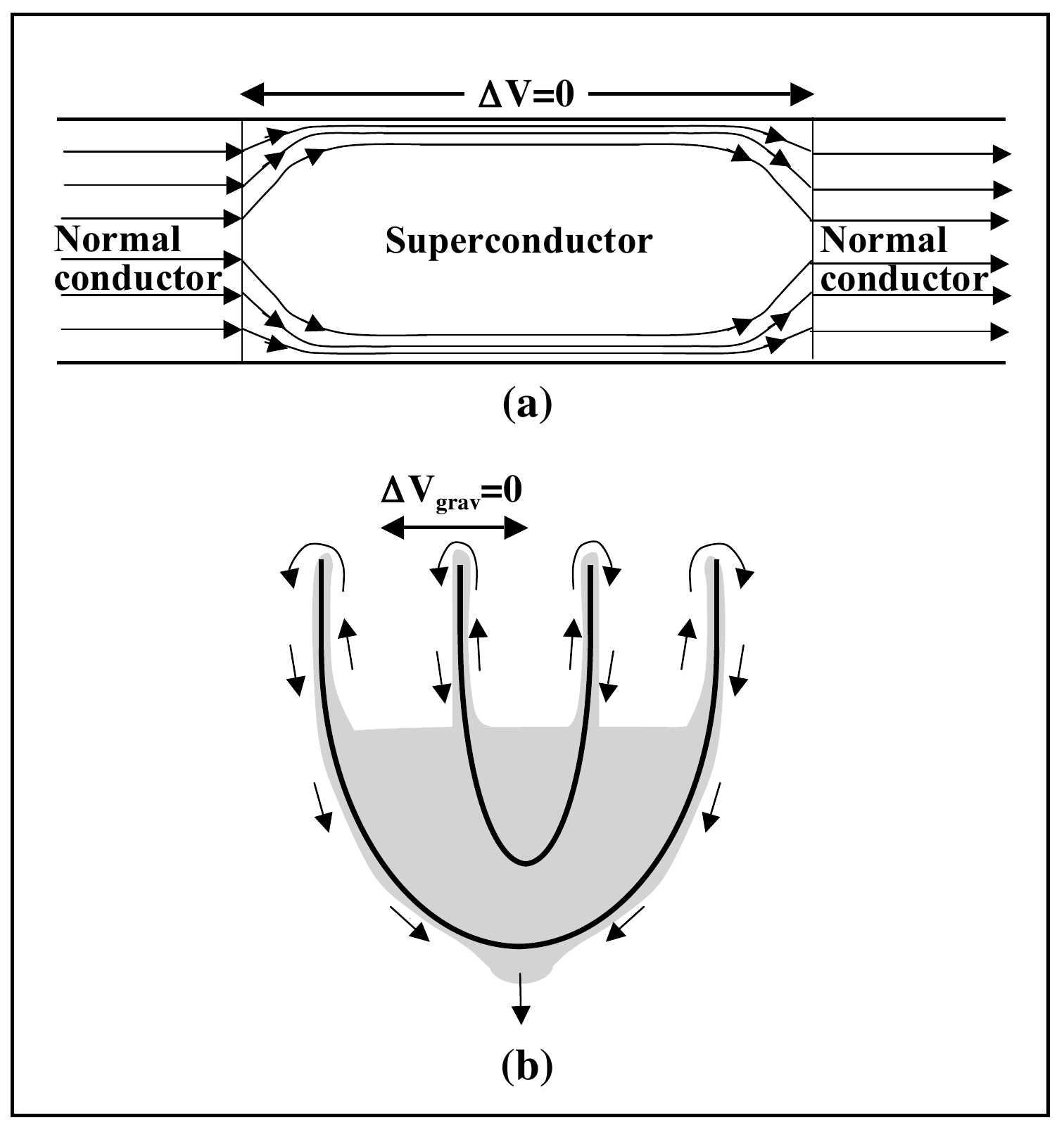}}
\caption {(a) Current flow in a superconducting wire which is fed by normal conducting leads.
  There is no electric potential difference between both ends
 of the superconductor ($\Delta V=0$). (b) Flow of superfluid $^4He$ along surfaces in double beaker experiment. 
There
 is flow of superfluid from the inner to the  outer beaker and from the outer beaker to the exterior, gradually emptying
 both beakers.  The levels in the inner and outer beaker are always  identical throughout this
 process, so there is no gravitational potential difference 
between them  ($\Delta V_{grav}=0$).  
}
\label{figure1}
\end{figure}

 Furthermore, Daunt and Mendelssohn\cite{daunt} as well as London\cite{londontransfer} and Bilj et al\cite{bilj} pointed out that the measured speed of $^4He$ in the films obeys the relation
 \beq
 v=\frac{\hbar}{2m_{He} d}
 \eeq
 where $d$ is the thickness of the film, typically $\sim 300 \AA$, giving a speed $v\sim 26 cm/s$. This relation can be interpreted as arising from Heisenberg's uncertainty
 principle for a particle confined to a linear dimension $d$. Similarly, the critical magnetic field for a superconductor is given by\cite{tinkham}
 \beq
 H_{c1}=-\frac{\hbar c }{4e\lambda_L^2}
 \eeq
and the critical velocity by
\beq
v=\frac{e}{m_e c} \lambda_L H_{c1}=\frac{\hbar}{4m_e \lambda_L}
\eeq 
 which can be interpreted as the speed of an electron confined to linear dimension $2\lambda_L$ arising from Heisenberg's uncertainty principle. 
 Mendelssohn argues\cite{mendel} that these speeds, Eqs. (6) and (8),  are the speeds of ``zero point diffusion'' of particles in the condensate, and that this explains why the transport rate is
 independent of external forces: the transport occurs because if at one end particles of the condensate are removed, zero point diffusion will give rise to flow in that direction. 
 He furthermore stresses that ``the momentum of frictionless transport is not dissipated because it is zero-point energy''.
 
 However, Mendelssohn's interpretation, even though it reveals very deep intuition, is not    internally consistent. Heisenberg's uncertainty principle predicts that the momentum associated with spatial confinement should be in the same direction of the coordinate that is confined.
 Instead, both in the superfluid and superconductor the transport with speeds given by Eqs. (6) and (8) is parallel to the surface, i.e. perpendicular to the direction of confinement. It is clear that Heisenberg's
 uncertainty principle is $not$ the explanation for superfluid film and superconducting current flow under zero potential difference. So what is it?
 
 Superconductors give us the answer. The London-Mendelssohn transfer speed for superconductors Eq. (8) is nothing other than the speed Eq. (3) of electrons in $2\lambda_L$ orbits giving rise to the spin current near
 the surface. The motion described by the speed Eq. (3) is $rotational$ (Fig. 7, left side). Thus we conclude that both superconductors and superfluid $^4He$ must possess 
 {\it rotational zero point motion} in their ground states.\cite{heliumsc}
 
 If the zero-point motion is rotational, it is easy to understand why spatial confinement in direction perpendicular to the surface gives rise to flow along the surface. Furthermore it is easy to understand the
 magnitude of the flow velocity, arising from quantization of angular momentum
 \beq
 L=mvd=\frac{\hbar}{2}
 \eeq
 for both Eq. (6) and Eq. (8). It is also easy to understand the origin of quantum pressure in these systems: the kinetic energy of rotational zero point motion decreases as the
 radius of the motion increases:
 \beq
 E_{kin}=\frac{L^2}{2MR^2}
 \eeq
 for particles of mass $M$ in orbits of radius $R$ with angular momentum $L$. Thus, a rotating   particle with fixed quantized angular momentum exerts quantum pressure to lower its kinetic
 energy by expanding its orbit, and does so in the transition to the superfluid or superconducting state. The expanded orbits overlap, hence phase coherence is required to avoid collisions of particles in
 different orbits, which is clearly a lower entropy state than when the phases are incoherent in the normal state, hence the transition will occur at sufficiently low temperatures where
 the energy decrease dominates over the entropy loss.
  
   \section{conclusion}
 
In summary, we conclude that in both superconductors and superfluid $^4He$ the transition to the superconducting or superfluid state is driven by quantum pressure originating in rotational zero point motion, 
i.e. the drive of a rotating system to lower its kinetic energy by  expansion. This explains a variety of properties of $^4He$ like the decrease in density below the superfluid transition, the shape
of the heat capacity curve versus temperature that gives the $\lambda-$transition its name, and the flow of Rollin films,
as well as the most fundamental property of superconductors, the Meissner effect.

We should point out that there have been several proposals in the literature that $^4He$ possesses macroscopic quantum zero point motion in the ground state\cite{zero1,zero2,zero3,zero4},
and that superconductors possess macroscopic zero point motion in the form 
of charge currents over domains\cite{curr1}. 
These workers arrived at these conclusions 
 through arguments different from ours.

 Finally, the facts that superconductors and superfluid $^4He$  are macroscopic quantum systems and they both display quantum pressure
originating in rotational zero point motion at the macroscopic level leads us to conclude that quite generally  $microscopic$ quantum systems, which also 
exhibit quantum pressure, {\it must acquire this quantum pressure through rotational zero point motion}. In other words, that the
origin of the ubiquitous quantum pressure is not Heisenberg's uncertainty principle as generally believed but instead  rotational zero point motion.
Since Schr\"{o}dinger's equation does not predict rotational zero point motion, this implies that
Schr\"{o}dinger's equation needs to be modified. The constant $\hbar$ in Schr\"{o}dinger's 
equation presumably represents the angular momentum of this ubiquitous rotational zero point motion
rather than the quantum of  action as in the conventional understanding of quantum mechanics.


\begin{references}
\bibitem{london12} F. London has written a two-volume book series entitled `Superfluids' (Wiley, New York),   Volume I 
(1950) on
superconductors and Volume II (1954) on superfluid $^4He$, emphasizing the common aspects of the phenomena.

   \bibitem{tilley} D.R. Tilley and J. Tilley, ``Superfluidity and Superconductivity'', 
          Institute of Physics Publishing, Bristol, 1990.
          
          
\bibitem{99} J.E. Hirsch and F. Marsiglio, Physica C {\bf 331}, 150 (2000). 
\bibitem{ceperley} D. Ceperley, Rev. Mod. Phys. {\bf 67}, 279 (1995).
\bibitem{holesc} J.E. Hirsch and F. Marsiglio, Phys. Rev. B {\bf 39}, 11515 (1989).
\bibitem{apparent} J.E. Hirsch, Physica C {\bf 199}, 305 (1992). 

\bibitem{londonbook}    F. London, Superfluids, Vol. II, John Wiley and Sons, Hoboken, New Jersey, 1954.
\bibitem{goldstein} L. Goldstein and J. Reekie, Phys. Rev. {\bf 98}, 857 (1955).
\bibitem{isopyc}  W.H. Keesom, ``Helium'', Elsevier Publ. Co., Amsterdam, 1942, p. 241.

\bibitem{simon} The key role of zero point   energy in liquid $^4He$ was pointed out early on by F. Simon, Nature {\bf 133}, 529 (1934) and references therein, and emphasized by F. London\cite{londonbook}.


\bibitem{website} See the website http://physics.ucsd.edu/~jorge/hole.html  for a complete list of references.
\bibitem{electronhole2} J.E. Hirsch, Phys. Rev. B {\bf 71}, 104522 (2005).
\bibitem{chargeexpulsion} J.E. Hirsch, Phys. Rev. B {\bf 68}, 184502 (2003). 
\bibitem{sm} J.E. Hirsch, Europhys. Lett. {\bf 81}, 67003 (2008)
\bibitem{meissner} J.E. Hirsch, Physica Scripta {\bf 85}, 035704 (2012). 
\bibitem{plasma} P.A. Davidson, ``An Introduction to Magnetohydrodynamics", Cambridge University Press, Cambridge, 2001.

\bibitem{japan} J.E. Hirsch, Physica C {\bf 470}, S955 (2010).
\bibitem{electrospin}  J.E. Hirsch,  Ann. Phys. (Berlin) {\bf 17}, 380 (2008). 

\bibitem{thexp1} J.J. Neumeier et al, Phys. Rev. B{\bf 72}, 220505 (2005).
\bibitem{thexp2} J.L. Olsen and H Rohrer, Helv. Phys. Acta {\bf 31}, 289 (1958).

 \bibitem{onnes} H. Kamerlingh Onnes, Leiden Comm. {\bf 159}, Trans. Faraday Soc. {\bf 18},
        No. 53 (1922).
       \bibitem{rollin} B.V. Rollin and F. Simon, Physica {\bf 6}, 219 (1939).
\bibitem{films}       J.G. Daunt and K. Mendelssohn, Nature {\bf 141}, 911 (1938); {\bf 142}, 475 (1938);
        {\bf 143}, 719 (1939); {\bf 157}, 839 (1946); Proc. Roy. Soc. London A{\bf 170}, 423 (1939); A{\bf 170}, 439 (1939).
\bibitem{beaker} J.G. Daunt and K. Mendelssohn, Nature {\bf 157}, 839 (1946).
\bibitem{emf} J.E. Hirsch,  J. Sup. Nov. Mag. {\bf 23}, 309 (2010).
\bibitem{mendel} K. Mendelssohn, Proc. Phys. Soc. London  {\bf 57}, 371 (1945).
         \bibitem{mendel2} K. Mendelssohn, ``Report of an International Conference on Fundamental Particles and Low Temperatures'', 
                 Physical Society of London, 1947, Vol. II, p. 35.
\bibitem{daunt} J.G. Daunt and K. Mendelssohn, Nature {\bf 150}, 604 (1942); Phys. Rev. {\bf 69}, 126 (1946).
\bibitem{londontransfer} F. London, Rev. Mod. Phys. {\bf 17}, 310 (1945).
\bibitem{bilj} A. Bilj, J. De Boer and A. Michels, Physica {\bf 8}, 655 (1941).
      \bibitem{tinkham} M. Tinkham, ``Introduction to Superconductivity'', 2nd ed., McGraw-Hill, New York, 1996.
      \bibitem{heliumsc} J.E. Hirsch, Mod. Phys. Lett. B {\bf 25}, 2219 (2011). 
      \bibitem{zero1} S.R. Shenoy and A.C. Biswas, Jour. Low Temp. Phys. {\bf 28}, 191 (1977).
          \bibitem{zero2} V.I. Yukalov, Physica {100A}, 431 (1980).
              \bibitem{zero3} Y.S. Jain, arxiv:1008.0240 (2010), arxiv:1011.1552 (2010), arxiv:1011.3190 (2010).
              \bibitem{zero4} V. A. Golovko, Physica A {\bf 246}, 275  (1997).
              \bibitem{curr1} F. Bloch, unpublished;  L. Landau, Phys. Zeits. der Sowjet. {\bf 4}, 43 (1933); J. Frenkel, Phys. Rev. {\bf 43}, 907 (1933); 
                 H.G. Smith and J.O. Wilhelm, Rev. Mod. Phys. {\bf 7}, 237 (935); 
   M. Born and K.C. Cheng, Nature {\bf 161}, 968 (1948);  W. Heisenberg, Z. fur Naturf. {\bf 2a}, 185 (1947); H. Koppe, Ann. der Physik (Leipzig) {\bf 1}, 405 (1947). 

                      
  \end{references}
\end{document}